\documentclass[sigconf]{acmart}

\usepackage{amsmath,amsfonts}
\usepackage{graphicx}
\usepackage{textcomp}
\usepackage{xcolor}
\usepackage{booktabs}
\usepackage{multirow}
\usepackage{textcomp}
\usepackage[flushleft]{threeparttable}
\usepackage{siunitx}
\usepackage{subcaption}

\usepackage{enumitem}
\usepackage{noindentafter}
\NoIndentAfterEnv{itemize}
\NoIndentAfterEnv{description}
\NoIndentAfterEnv{enumerate}
\setlist{
	leftmargin=14pt,
	itemsep=1pt,
}
\setdescription{
    font=\normalfont\itshape
}
\renewcommand{\descriptionlabel}[1]{%
  \hspace\labelsep \upshape #1:%
}

\copyrightyear{2024} 
\acmYear{2024} 
\setcopyright{rightsretained} 
\acmConference[MSR '24]{21st International Conference on Mining Software Repositories}{April 15--16, 2024}{Lisbon, Portugal}
\acmBooktitle{21st International Conference on Mining Software Repositories (MSR '24), April 15--16, 2024, Lisbon, Portugal}\acmDOI{10.1145/3643991.3644897}
\acmISBN{979-8-4007-0587-8/24/04}

\NewDocumentCommand\code{+m}{{\ttfamily #1}}

\NewDocumentCommand\RQ{+m}{{\bf RQ$_{#1}$}}

\NewDocumentCommand{\mypm}{mm}{#1 {\color{gray}$\pm$ #2}}

\begin{document}

\title{On the Effectiveness of Machine Learning-based Call Graph Pruning: An Empirical Study}


\author{Amir M. Mir}
\affiliation{%
  \institution{Delft University of Technology}
  \city{Delft}
  \country{The Netherlands}
}
\email{s.a.m.mir@tudelft.nl}

\author{Mehdi Keshani}
\affiliation{%
  \institution{Delft University of Technology}
  \city{Delft}
  \country{The Netherlands}
}
\email{m.keshani@tudelft.nl}

\author{Sebastian Proksch}
\affiliation{%
  \institution{Delft University of Technology}
  \city{Delft}
  \country{The Netherlands}
}
\email{s.proksch@tudelft.nl}

\renewcommand{\shortauthors}{Mir et al.}

\begin{abstract}
Static call graph (CG) construction often over-approximates call relations, leading to sound, but imprecise results.
Recent research has explored machine learning (ML)-based CG pruning as a means to enhance precision by eliminating false edges.
However, current methods suffer from a limited evaluation dataset, imbalanced training data, and reduced recall, which affects practical downstream analyses.
Prior results were also not compared with advanced static CG construction techniques yet.
This study tackles these issues.
We introduce the NYXCorpus, a dataset of real-world Java programs with high test coverage and we collect traces from test executions and build a ground truth of dynamic CGs.
We leverage these CGs to explore conservative pruning strategies during the training and inference of ML-based CG pruners.
We conduct a comparative analysis of static CGs generated using zero control flow analysis (0-CFA) and those produced by a context-sensitive 1-CFA algorithm, evaluating both with and without pruning. 
We find that CG pruning is a difficult task for real-world Java projects and substantial improvements in the CG precision (+25\%) meet reduced recall (-9\%).
However, our experiments show promising results: even when we favor recall over precision by using an F2 metric in our experiments, we can show that pruned CGs have comparable quality to a context-sensitive 1-CFA analysis while being computationally less demanding.
Resulting CGs are much smaller (69\%), and substantially faster (3.5x speed-up), with virtually unchanged results in our downstream analysis.

\end{abstract}





\maketitle

\section{Introduction}
Call graphs (CG) represent function invocations within programs~\cite{ryder1979constructing, callahan1990constructing}.
Their construction is a crucial component of static program analysis, like security analysis, dead code identification, performance profiling, and more.
An ideal CG would be both \emph{sound}, i.e., not missing any legitimate function call, and \emph{precise}, i.e., not containing unnecessary function calls.
However, constructing a sound and precise CG is challenging even for small programs~\cite{ali2012application}.
In practice, static CG construction will over-approximate the call relations to boost soundness at the cost of precision: popular tools like WALA~\cite{fink2012wala} or Petablox~\cite{mangal2015user} create imprecise CGs with up to 76\% false edges~\cite{utture2022striking}.
To address this imprecision, previous work~\cite{bravenboer2009strictly, mangal2015user, tan2016making} have enhanced pointer analysis, which builds the backbone of numerous CG construction algorithms, by improving the context-sensitivity or flow-sensitivity.
Unfortunately, a flawless pointer analysis is principally infeasible~\cite{rice1953classes}, and pointer analyses often require a tradeoff between scalability and precision~\cite{li2018scalability}.
For instance, WALA's context-sensitive analysis only reduces the false positive rate by 8.6\% compared to a context-insensitive analysis, despite significantly slowing down performance~\cite{utture2022striking}.

Recent work has introduced Machine Learning (ML)-based call graph pruning approaches to improve the precision of call graphs by pruning false edges in call graphs as a post-processing step.
Techniques like CGPruner~\cite{utture2022striking} and AutoPruner~\cite{le2022autopruner} learn from dynamic traces that are collected in actual program executions to identify unnecessary edges in a static CG.
CGPruner only leverages features of the CG structure, while AutoPruner combines structural features with automatically extracted semantic features from the source code that are encoded with the code language model (CLM), CodeBERT~\cite{feng2020codebert}.
Although these previous approaches show intriguing results, they suffer from several limitations:
(1) Both have been trained and evaluated on the NJR-1 dataset~\cite{palsberg2018njr}, which lacks real-world projects and suffers from a notoriously low branch coverage (68\%).
(2)
The over-approximation of static call graph construction results in many unnecessary edges~\cite{reif2021novel} while dynamic CGs contain much fewer edges.
As a result, the training and evaluation of the ML models have to deal with a highly imbalanced dataset.
(3) After CG pruning, the recall drops substantially by more than 25\%~\cite{le2022autopruner}, which makes the pruned CGs impractical for client analyses, especially for security-focused applications.
(4) The previous work used a 0-CFA algorithm to generate static CGs, which is context-insensitive and less precise.
It is unclear how advanced, context-sensitive CG algorithms like $k$-CFA algorithms~\cite{shivers1991control} perform in comparison.

In this paper, we will address these issues by
(1)~introducing a meta dataset, NYXCorpus, which includes the existing datasets NJR-1~\cite{palsberg2018njr}, and XCorpus~\cite{dietrich2017xcorpus}.
We also added YCorpus, which is based on another dataset of projects with a high test coverage of 88\%~\cite{khatami2023state}.
In contrast to NJR-1, both XCorpus and YCorpus contain real-world projects.
We combined these three datasets and generated dynamic traces through test execution to create a unified benchmark.
(2)~To address the second and third issues, we explore a conservative pruning strategy during the learning phase and different confidence levels for the inference of ML models to prune CG edges. These two strategies help to deal with an imbalanced dataset and mitigate the recall drop after pruning.
(3)~In addition to 0-CFA, we also use 1-CFA to generate static CGs and compare both algorithms with and without pruning in terms of quality and scalability.

\NewDocumentCommand{\rqOne}{}{How do ML-based CG pruning models generally perform at a CG pruning task?}
\NewDocumentCommand{\rqTwo}{}{Can conservative training/pruning strategies improve the results?}
\NewDocumentCommand{\rqThree}{}{How do context-sensitive CG generators compare in terms of quality and scalability?}
\NewDocumentCommand{\rqFour}{}{Is CG pruning practical for a security application like vulnerability analysis?}

We will answer the following research questions to investigate the impact of these three improvements:

\begin{description}
\item[\RQ{1}] \rqOne
\item[\RQ{2}] \rqTwo
\item[\RQ{3}] \rqThree
\item[\RQ{4}] \rqFour
\end{description}

Our main results show that CG pruning is difficult on real-world Java projects.
Although ML-based call graph pruning techniques are effective at boosting the precision of static CGs, the recall drops as a result.
Our experiments report F2 values to prioritize recall over precision, but even then the tradeoff is in favor of the ML pruners.
Pruned CGs have comparable quality to a context-sensitive 1-CFA analysis, while their creation is computationally less demanding.

Our pruners can be configured by incorporating weights in the learning process or confidence levels when pruning to control the resulting precision and recall.
Our experiments show that a well-configured pruner can improve the quality of a 0-CFA CG more than running a more advanced 1-CFA analysis would.
We will show that both have a similar execution time, but that the pruned CG has a higher quality and is smaller.
We use the resulting CGs in a use case analysis of a security-focused application, in which we investigate the reachability of vulnerable methods.
We can show that analyses using pruned CGs generate very minimal false negatives (less than 2\%) while benefiting from a faster analysis time of up to 5 times due to the reduced size of pruned CGs. 

\noindent
Overall, this paper makes the following main contributions.
\begin{itemize}
\item We created a new benchmark dataset, NYXCorpus, from pre-existing datasets and tailored it to the call graph pruning task. It has Java programs of various sizes including real-world ones.
\item We adapt existing ML models to support weighted training and customizable pruning through confidence levels.
\item We present an empirical study on the effectiveness of ML-based call graph pruning, which studies current issues, proposes solutions, and evaluates their effects.
\end{itemize}

The rest of this paper is organized as follows. We describe related work in section~\ref{sec:related-work}. We explain our research methodology in section~\ref{sec:approach}. The evaluation setup for this study is described in section~\ref{sec:eval-setup}. We present the obtained empirical results in section~\ref{sec:eval}. We discuss the implications of the obtained results in section~\ref{sec:discuss}. We describe threats to validity and limitations in section~\ref{sec:ttv}. Finally, we conclude our empirical study in section~\ref{sec:summary}.

\section{Related Work}\label{sec:related-work}
\paragraph{Call Graph Construction}
Call graph construction has been widely studied. ML-based call graph pruner
does not utilize run-time information and hence it falls into the category of
static approaches~\cite{murphy1998empirical, reif2019judge, sui2020recall} for constructing call graphs. Approaches that use dynamic analysis~\cite{xie2002empirical, hejderup2018software} result in fewer false positives and higher precision, but they are less scalable.

Also, research has been conducted to enhance the precision of call graphs. Lhotak~\cite{lhotak2007comparing} created an interactive tool to help understand the root cause of discrepancies between different static and dynamic analysis tools. Sawin and Rountev~\cite{sawin2011assumption} proposed specific heuristics to manage dynamic features like reflection, dynamic class loading, and native method calls in Java. This approach improved the precision of the Class Hierarchy Analysis (CHA) algorithm~\cite{dean1995optimization} while maintaining decent recall levels. Moreover, Zhang and Ryder~\cite{zhang2007automatic} worked on generating precise application-only call graphs by distinguishing false-positive edges between the standard library and the application. 
Similar to the described work, ML-based CG pruners~\cite{utture2022striking, le2022autopruner} aim to improve CG precision as a data-driven post-processing approach by removing false edges.

\paragraph{Call graph comparison}
Xie and Notkin~\cite{xie2002empirical} quantitatively and qualitatively compared dynamic and static call graphs from two Java micro-benchmarks. They found that static call graphs tend to be conservative but imprecise due to computational complexity. Dynamic call graphs, on the other hand, are more straightforward and reflect the actual invocations.
Lhotak~\cite{lhotak2007comparing} presented a technique to find the root causes of call graph differences and the PROBE framework. PROBE facilitates comparing dynamic and static call graphs to identify sources of imprecision.
In this study, we compare pruned static call graphs for Java programs to their dynamic call graphs, and we analyze the differences between them in terms of precision and soundness. 

\paragraph{Machine learning-based call graph pruning}
As of this writing, there are currently two ML-based call graph pruning models, CGPruner~\cite{utture2022striking} and AutoPruner~\cite{le2022autopruner}. Utture et al.~\cite{utture2022striking} introduced an ML-based technique called CGPruner, with the goal of reducing the false-positive rate of static analysis tools, making them more attractive to developers. CGPruner prunes the static call graph, which is at the core of many static analyses, by removing false-positive edges while retaining true edges. The technique achieves this balance using an ahead-of-time learning process involving executing static and dynamic call-graph constructors. The dynamic call graphs were only used during a training phase on a training set of programs. CGPruner was shown to significantly decrease the false-positive rate, in one case, from 73\% to 23\%.

CGPruner does not consider source code semantics. To address this limitation, Le-Cong et al. proposed AutoPruner~\cite{le2022autopruner} to prune false positives in call graphs by leveraging both structural and statistical semantic information. The semantic features extracted from the caller and callee functions' source code. Specifically, AutoPruner uses CodeBERT~\cite{feng2020codebert}, a pre-trained Transformer model~\cite{vaswani2017attention} for code, fine-tuning it to capture semantic features for each edge and combines them with handcrafted structural features, and employs a neural classifier to classify each edge as true or false-positive.

\paragraph{Machine Learning for Software Engineering}
In recent years, the application of machine learning for software engineering has been a hot topic of research~\cite{allamanis2018survey, sharma2021survey}. ML models have been used to perform various tasks, such as code completion, code summarization, defect prediction, code classification, and code translation tasks. Recently, large-scale code language models (CLMs)~\cite{zan2022neural} such as CodeBERT~\cite{feng2020codebert} and CodeT5~\cite{wang2021codet5} have achieved state-of-the-art performance on numerous SE tasks mentioned above. In general, ML can offer opportunities to improve or automate several aspects of the traditional software development process. The scale of software artifact data, automated feature engineering provided by ML techniques, robustness and scalability of optimization techniques, and transferability of traditional ML applications to SE artifacts all indicate the potential of ML to improve the traditional software development process. This research area is called Machine Learning for Software Engineering (ML4SE).

\section{Approach}\label{sec:approach}
\begin{figure}
 \centering
 \includegraphics[width=.8\columnwidth]{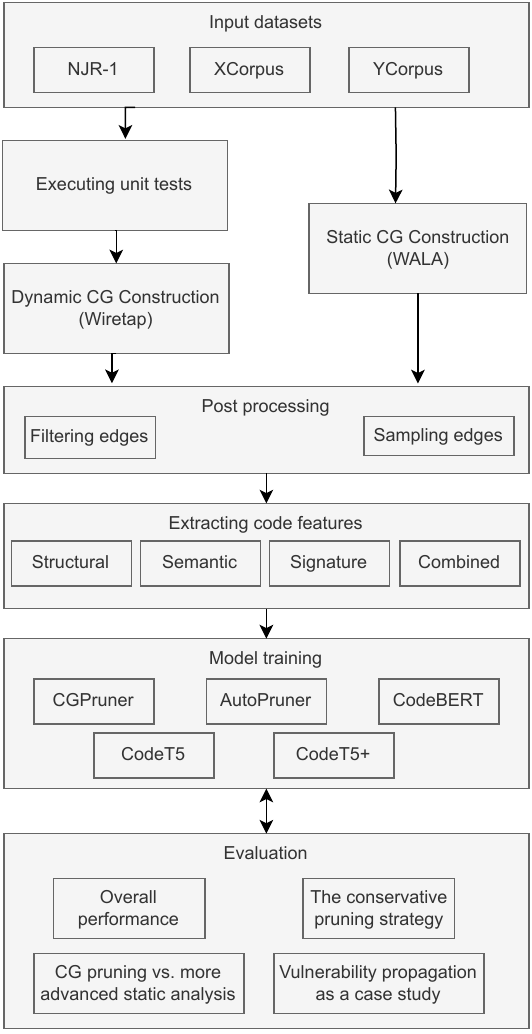}
 \caption{Overview of our approach used in this study}
 \label{fig:overall}
\end{figure}

In this section, we first define the research problem under study. Then, we introduce the various ingredients of our research methodology: the \emph{datasets} that we use in our experiments, a description of the \emph{call-graph generation} (both static and dynamic), an explanation of ML models used in previous works~\cite{utture2022striking, le2022autopruner}, and recent suitable code language models for this task; lastly, we describe the different code features that we use for training call graph pruners. Figure~\ref{fig:overall} shows an overview of our research methodology used in this empirical study. Overall, our proposed methodology consists of three datasets, static/dynamic CG construction, post-processing like filtering/sampling edges, training of ML models, and empirical evaluation. All these steps are presented later in the paper.

\subsection{Problem definition}
This paper studies CG pruning, which takes a static call graph $\mathbb{G}$ as initial input. A CG is a directed graph created using a static analysis tool.
The vertices $V$ of the graph represent defined functions, which are identified by a function signature (name, parameters, return type).
The edges $E$ represent calls from one function to another.
Each edge within $E$ is defined as a tuple, that consists of the calling function (caller), the function being called (callee), and the site within the caller where the call is made (offset).

The output $\mathbb{G}^\prime$ is a refined version of the original CG, where $\mathbb{G}^\prime = (V^\prime, E^\prime)$, $V^\prime=V$, and $E^\prime$ is a subset of $E$.
The reduction is achieved through a binary classifier, $C$, which is designed to decide per edge $e \in E$, whether the edge should be copied to $\mathbb{G}^\prime$ or pruned.
Our validation is based on dynamic CGs that we construct from traces of actual program and test executions and that we use to validate the pruned call graph $\mathbb{G}^\prime$.

\subsection{Datasets}\label{subsec:datasets}
In this section, we describe the three datasets of our study, that we use to train and evaluate the ML-based CG pruning models.

\paragraph{NJR-1}
Normalized Java Resource (NJR)~\cite{palsberg2018njr} is an infrastructure to leverage the potential of Big Code. The normalization enables searchability, scriptability, and reproducibility. The NJR comprises 100,000 executable Java programs, a set of pre-existing tools, which facilitate the development of novel research tools. For evaluating the ML-based call graph pruning models, we use a subset of the NJR1 dataset, created by the work of Utture et al.~\cite{utture2022striking}. The subset contains 141 programs from the NJR-1 benchmark suite, of which 100 programs are used for training the models and 41 programs for evaluation. The selection of 141 programs from NJR-1 programs was based on criteria such as each program having at least 1,000 methods and 2,000 static call graph edges as per Wala, executing a minimum of 100 distinct methods during runtime, and exhibiting high coverage, i.e., executing a large portion of the methods that can be reached from the main method (with an average coverage of 68\%). On average, each selected program comprises around 560,000 lines of code, excluding the standard library~\cite{utture2022striking}. 

\paragraph{XCorpus}
The XCorpus dataset~\cite{dietrich2017xcorpus} contains a set of 76 executable Java programs, which includes 70 from the Qualitas Corpus~\cite{tempero2010qualitas}. This corpus combines both built-in and generated test cases, offering better branch coverage than the DaCapo benchmark~\cite{blackburn2006dacapo}. While the DaCapo benchmark and Qualitas Corpus are curated datasets for benchmarking and static analysis, respectively, XCorpus combines the strengths of both—being executable (like DaCapo) and diverse and extensive (like Qualitas Corpus). Such a dataset is useful for research on program analysis, studies combining static and dynamic analyses, or studies on program transformations that evaluate impact through program execution pre- and post-transformation. The average coverage for XCorpus' programs is moderate, with 62.35\% for built-in and 60.25\% for
generated tests by Evosuite~\cite{fraser2011evosuite}. On average, each program has an average of around 36K lines of code. XCorpus has been used in the empirical studies on the soundness of Java call graphs~\cite{sui2020recall, reif2019judge}.

\paragraph{YCorpus}
For this work, we created a new dataset, namely, YCorpus, based on an existing dataset used in a recent empirical study by Khatami and Zaidman~\cite{khatami2023state}, which investigates state-of-the-practice in quality assurance in Java-based open source software development. Specifically, they have studied 1,454 popular Java projects on GitHub with more than 100 stars. Given this, we selected 40 Java projects with the criteria that each project has higher than 80\% test coverage. These 40 projects have 88\% test coverage on average, which is substantially higher than that of XCorpus and NJR-1.
Also, YCorpus contains real-world Java projects such as Apache Commons IO, AssertJ, and MyBatis 3, and each project has an average of around 50K lines of code.
Both XCorpus and YCorpus have been reduced to the programs that we could build and for which we were able to construct both static and dynamic CGs.

\paragraph{NYXCorpus}
In the remainder of the paper, we will refer to NYXCorpus as a dataset to indicate that we have based our experiments on the joined data of all three corpora.

\paragraph{Source Code Recovery}
The original NJR-1 dataset lacks source code for the dependencies of its programs.
However, call graphs contain nodes/methods related to dependencies and code language models need source code to learn the call graph pruning task.
We have identified dependencies in the NJR-1 dataset, located their respective repositories (often on platforms like GitHub), and downloaded the necessary source code to extend the dataset with the original code, including comments, for deeper code understanding.
For XCorpus and YCorpus, we downloaded sources JARs for their programs via the Maven or Ant command-line tools.

\subsection{Call-Graph Generation}\label{subsec:CGG}
This subsection describes our dynamic and static CG generation and explains our filter and sampling criteria for program edges.

\paragraph{Dynamic Call Graphs} 
To evaluate (pruned) static call graphs, we need to establish an "oracle", a known ground truth that represents actual program behavior. In this context, the oracle refers to vertices in a call graph which represents methods. These methods are recognized using a mix of the class name where the method is defined, the method's name, and a descriptor, as per the Java language specification~\cite{lindholm2021java}, to account for overloading. The edges in call graphs are formed by pairs of source and target methods.
To obtain such an oracle, we utilize unit tests that are commonly available and can be an effective way to initiate program executions. In fact, built-in test cases offer a unique insight as they represent the \textit{intended} behavior of a program, mirroring the experience an end-user might have when using the software in a real-world setting~\cite{sui2020recall}.

To collect the method calls of a program, we have instrumented it via Wiretap~\cite{kalhauge2018sound}, a tool to trace information from a running Java program. Specifically, we wrote a recorder to insert probes at Java method entries and exits to record call relationships.
We then ran all available unit tests to gather execution paths, creating dynamic call graphs for the actual execution paths, serving as an oracle or "ground truth" for evaluating ML-based call graph models.
We use this \emph{dynamic} data to train a model for detecting and pruning irrelevant or infeasible paths in \emph{static} CGs.

\paragraph{Static Call Graphs}
We employ the WALA framework~\cite{fink2012wala} to generate static CGs with the \emph{context-insensitive} 0-CFA (Zeroth-order Control Flow Analysis)~\cite{shivers1991control}, and the \emph{context-sensitive} 1-CFA algorithm.
For this study, we chose WALA over alternatives like DOOP~\cite{smaragdakis2021doop} and Soot~\cite{vallee2010soot} as it has better support for Java language features such as lambda expressions and, as of this writing, it supports Java bytecode up to JDK 17~\cite{wala}.
We follow prior research work~\cite{utture2022striking, le2022autopruner} and do not use WALA's handler for Java reflection, which can potentially miss some execution paths that involve reflective calls.
In short, given a Java program, we perform the following steps to construct a static CG using the WALA framework:

\begin{itemize}
    \item We consider each project's main JAR file as the application scope and its transitive dependencies as the extension scope.
    \item We perform a \emph{Class Hierarchy Analysis}~\cite{dean1995optimization}, which involves constructing the class inheritance hierarchy to facilitate the resolution of method call targets.
    \item All non-private methods within all public classes are used as entry points for WALA's call graph builder.
    \item The obtained entry points and the CHA structure are used to construct 0/1-CFA static CGs.
\end{itemize}

\paragraph{Filtering edges} Considering the call graph pruning problem, we are interested in call graph edges related to the application itself and its dependencies.
We follow previous work and opt for removing edges to/from the Java standard library as its enormous size would dominate the dataset and skew the evaluation~\cite{utture2022striking}.
Specifically, we remove all call edges that start with the following prefixes: 
\code{java/}, \code{javax/}, \code{sun/}, \code{com/sun/}, \code{jdk/}.

\paragraph{Large Programs} Utture et al.~\cite{utture2022striking} observed that a few programs in the NJR-1 dataset have a very large number of call-graph edges (over 20K), and they randomly sampled 20K edges from the edge sets of those programs. Following this, we also randomly sampled 20K edges from the edge sets of 5 programs in the XCorpus and YCorpus datasets to alleviate the skewness in the dataset distribution. This also prevents bias towards large programs when training a model/classifier. Also, we do not remove or sample edges where they exist in both dynamic and static call graphs, as fewer of these true edges exist.
A removal of true edges would harm the performance of the ML models at retaining true edges, i.e., recall.

\subsection{Call-Graph Pruning Models}\label{sec:models}
In this subsection, we describe several machine learning techniques, including code language models, which we extend and employ for our CG pruning task.

\paragraph{Random Forest~\cite{ho1995random}} An ensemble learning method, constructs decision trees on bootstrapped datasets using Bagging~\cite{breiman1996bagging}, considering a random feature subset at each node. Predictions are derived from majority voting for classification or averaging for regression tasks. The algorithm is versatile and adept at handling numerous inputs, missing values, and errors in unbalanced datasets. However, it can be a "black box" model and may overfit noisy datasets.

\paragraph{CodeBERT~\cite{feng2020codebert}} A bimodal pre-trained model for programming language (PL) and natural language (NL) tasks, leverages a Trans\-former-based architecture~\cite{vaswani2017attention} and a hybrid objective function inclusive of replaced token detection during pre-training. It utilizes both bimodal and unimodal data, helping to learn better generators. Trained on CodeSearchNet~\cite{husain2019codesearchnet}, which contains GitHub repositories in six languages, it is similar to multilingual BERT without explicit language markers. Empirical results show that fined-tuned CodeBERT has superior performance on natural language code search and code documentation generation tasks. Without parameter fine-tuning, zero-shot setting tests also indicate the superiority of RoBERTa~\cite{liu2019roberta}, suggesting its effective learning and application in NL-PL tasks.

\paragraph{CodeT5~\cite{wang2021codet5}} Built on the T5 architecture~\cite{raffel2020exploring}, utilizes denoising sequence-to-sequence pre-training for both understanding and generation tasks in natural language. A novel identifier-aware pre-training task is introduced for better leveraging code semantics. Similar to CodeBERT, it is pre-trained on CodeSearchNet~\cite{husain2019codesearchnet} data and additional data from open-source GitHub  C/C\# repositories. It is fine-tuned on most CodeXGLUE benchmark tasks and supports multi-task learning. Experimental results reveal that CodeT5 outperforms CodeBERT on various tasks, demonstrating enhanced capture of semantic information from code.

\paragraph{CodeT5+~\cite{wang2023codet5+}} An adaptable family of encoder-decoder Large Language Models (LLMs) designed for code tasks, combining different pre-training objectives, including span denoising, contrastive learning, text-code matching, and causal language modeling, for flexible applications in various modes. Initiated with frozen off-the-shelf LLMs~\cite{nijkamp2022codegen}, it circumvents training from scratch, promoting efficient scaling. Upon evaluating 20+ code-related benchmarks, CodeT5+ exhibits superior performance in tasks including code generation, completion, and text-to-code retrieval.

\subsection{Code Features}\label{sec:feat}
We use the following features or code representations to train our ML-based CG pruning models.
The intuition behind all features is to provide information about the usefulness of an edge.


\paragraph{Structural}
Utture et al. \cite{utture2022striking} engineered a set of structural features encapsulating vital contextual and semantic call edge details, adhering to three criteria: linear-time computational complexity, interpretability/generalizability, and black-box nature. The proposed feature set is a combination of local and global information extracted from static call graphs (more info in~\cite{utture2022striking}). The structural features, $f_{struct}$, build a $k_{s}$-dimensional vector ($k_{s}=11$):
\begin{equation}
\mathbf{f}_{struct} = [x_{1}^{struct}, x_{2}^{struct}, ..., x_{k_{s}}^{struct}]
\end{equation}


\paragraph{Semantic}
Semantic features are extracted from the source code of the caller and callee functions, which is also used in the work of Le-Cong et al.~\cite{le2022autopruner}. Unlike hand-crafted structural features, semantic features are automatically learned by using code language models. They can generate a high-dimensional vector that captures the statistical relationships between caller and callee functions. Thus, each edge in the call graph has an associated embedding that represents the semantic relationship between the caller and the callee. Conceptually, semantic features are represented as follows:
\begin{equation}
\small
\texttt{[CLS]} \langle \texttt{caller's source} \rangle \texttt{[SEP]} \langle \texttt{callee's source} \rangle \texttt{[EOS]}
\end{equation}
The semantic features, $f_{sem}$, are represented as a $k_{c}$-dimensional vector ($k_{c} = 768$), which are the output embeddings of a code language model such as CodeT5.
\begin{equation}
\mathbf{f}_{sem} = [x_{1}^{sem}, x_{2}^{sem}, ..., x_{k_{c}}^{sem}]
\end{equation}

\paragraph{Signature-based}
AutoPruner~\cite{le2022autopruner} extracts features from caller and callee method signatures to supplement CG nodes without source code~\cite{apimp}, namely, \emph{class \& method name}, \emph{parameters}, and \emph{return types}.
This code feature provides minimal code context but is helpful when source code is unavailable.
Signature-based features, $f_{sig}$, are represented as a $k_{c}$-dimensional vector:
\begin{equation}
\mathbf{f}_{sig} = [x_{1}^{sig}, x_{2}^{sig}, ..., x_{k_{c}}^{sig}]
\end{equation}

\paragraph{Combined}
Le-Cong et al.~\cite{le2022autopruner} proposed a combined feature set, which takes advantage of both structural and semantic features, to prune call graph edges effectively. It has empirically shown that AutoPruner~\cite{le2022autopruner}, CodeBERT, with the combined feature set, outperforms a RandomForest model trained on structural features. The combined features, $f_{comb}$, are represented as the concatenation of two vectors $\mathbf{x}_{sem}$ and $\mathbf{x}_{struct}$:
\begin{equation}
\mathbf{f}_{comb} = \mathbf{x}_{struct} \oplus \mathbf{x}_{sem}
\end{equation}

\subsection{Model Training}
We fine-tune our code language models 
for two epochs. Specifically, only the encoder module of the CLMs is fine-tuned, which generates embedding for code features mentioned in subsection~\ref{sec:feat}. To speed up training, we utilized mixed precision training with a floating point precision of 16-bit, which effectively reduces the GPU memory consumption without sacrificing the model's performance. We used an initial learning rate of $1 \times 10^{-5}$, which was found to be effective for training such models without causing instability in the learning process~\cite{le2022autopruner}.
We use cross-entropy loss as the loss function, which is suitable for binary classification problems:
\begin{equation}\label{eq:cross-entropy}
L(y, \hat{y}) = -\frac{1}{N} \sum_{i=1}^{N} [w_{1} * y_{i} * log(\hat{y}_{i}) + w_{2} * (1-y_{i}) * log(1-\hat{y}_{i})]
\end{equation}
Where $L(y, \hat{y})$ is the loss function comparing the true labels $y$ and the predicted labels $\hat{y}$. $N$ is the total number of samples. $w_{1}$ and $w_{2}$ are the weights associated with the positive and negative classes, respectively. This allows us to define pruning strategies like "conservative" by giving a higher weight to the positive class. The conservative strategy prioritizes high recall by maintaining as many edges as possible and only prunes when certain. We study the effect of these weights on pruning call graphs in RQ2.

We used the AdamW optimizer~\cite{loshchilov2017decoupled}, a variant of the Adam optimizer that corrects its weight decay regularization, boosting generalization and controlling over-fitting. A dropout rate of 0.25 was applied to the models' encoder output to prevent overfitting further~\cite{srivastava2014dropout}. Also, we adopted a linear scheduling policy~\cite{goyal2017accurate} with a warmup phase of 100 steps. This method gradually ramps up the learning rate from zero to the specified maximum ($1 \times 10^{-5}$ in this case) during the warmup phase to avoid large gradient updates early in training, thus aiding in better convergence.

\subsection{Model Inference}\label{subsec:model-inf}
Given a fine-tuned code language model, we prune CG edges as follows. Let \( \mathbf{x} \) be an input to the CLM and the linear neural network (NN) produces raw scores for the two classes, \( z_0 \) and \( z_1 \). Then, the softmax function converts these raw scores into probabilities.
\begin{equation}
p(y = i | \mathbf{x}) = \frac{e^{z_i}}{e^{z_0} + e^{z_1}}
\end{equation}
Where \( i \) is the class label which can take values 0 or 1 and $p(y = i | \mathbf{x})$ is the probability of class $i$. Finally, we use a decision function with a threshold (e.g., \( \tau = 0.5 \)) to decide the predicted class as follows.
\begin{equation}\label{eq:decision_func}
\hat{y} = 
\begin{cases} 
1 & \text{if } p(y = 1 | \mathbf{x}) > \tau \\
0 & \text{otherwise}
\end{cases}
\end{equation}
In many cases with NNs having two output neurons, since the probabilities produced by the softmax function for both classes sum to 1, a threshold of 0.5 is commonly used. If \( p(y = 1 | \mathbf{x}) > 0.5 \), it automatically implies \( p(y = 0 | \mathbf{x}) < 0.5 \) and vice versa.

\begin{table}
\centering
\caption{Stats for the datasets used in evaluation}
\label{tab:data-stats}
\small
\begin{threeparttable}
\begin{tabular}{@{}l rr c rr r@{}}
\toprule
\multirow{2}{*}{\textbf{Dataset}} & \multicolumn{2}{c}{\textbf{Num. Edges}} && \multicolumn{2}{c}{\textbf{Num. Tokens}} & \multirow{2}{*}{\textbf{P/R Ratio}}\tnote{1} \\
\cmidrule{2-3} \cmidrule{5-6}
 & \textbf{Train} & \textbf{Test} && \textbf{Train} & \textbf{Test} & \\
\midrule
NJR-1 & 859K & 405K && 262M & 124M & 18.7 \\
XCorpus & 269K & 57K && 22M & 8M & 2.4 \\
YCorpus & 242K & 72K && 35M & 9M & 3.7 \\
NYXCorpus & 1.37M & 534K && 319M & 141M & 7.6 \\
\bottomrule
\end{tabular}
\begin{tablenotes}
\item[1] Ratio of to-be-pruned and to-be-retained edges.
\end{tablenotes}
\end{threeparttable}
\end{table}

\section{Evaluation Setup}\label{sec:eval-setup}
In this section, we explain the evaluation metrics for evaluating (pruned) call graphs, the implementation details, model training, and the characteristics of the datasets used in this study. 

\paragraph{Evaluation Metrics}\label{sec:eval-metrics}
Similar to the previous work~\cite{utture2022striking, le2022autopruner}, to assess the accuracy of a static call graph, we use the common evaluation metrics, precision and recall. We denote the edge set generated by a static call-graph constructor as $E_S$, and the edge set created by Wiretap as $E_D$. The proportion of incorrect identifications is represented by (1-Precision). To obtain the average precision and recall values for the complete test set, we calculate the mean precision and recall values of individual programs and the $F_{\beta}$ measure.
\begin{align*}
\text{Precision} = \frac{|E_S \cap E_D|}{|E_S|} \quad \quad \text{Recall} = \frac{|E_S \cap E_D|}{|E_D|}
\end{align*}
\begin{equation*}
F_\beta = \frac{(1 + \beta^2) \times Precision \times Recall}{\beta^2 \times Precision + Recall}
\end{equation*}

where $\beta$ determines the weight of precision in the combined score. $\beta < 1$ gives more weight to precision, while $\beta > 1$ favors recall. We report $F_1$ and $F_2$ in our evaluation.

\paragraph{Implementation and Environmental setup}
To parse and extract Java methods' source code, we utilized a Java parser~\cite{javap}. We used PyTorch 2.0~\cite{paszke2019pytorch, pt2} with PyTorch Lightning 2.0~\cite{ptl2} to train and evaluate code language models described in section~\ref{sec:models}. We used the pre-trained code language models from HuggingFace's transformers library~\cite{wolf2019huggingface}. To implement the CGPruner model, i.e., RandomForest, we employed the scikit-learn library~\cite{pedregosa2011scikit}. We used NetworKit~\cite{angriman2022}, a toolkit for large-scale network analysis with optimized algorithms to process graph data and extract structural features. We also used JGraphT~\cite{jgrapht} in Java to do graph traversals and reachability analysis. We performed all the experiments on a Linux workstation (Ubuntu 22.04 LTS) with Intel Core i9 13900KS@6GHz, an RTX 4090 24GB, and 2x48GB (96GB) DDR5 RAM.

\begin{table*}
\centering
\caption{Comparison of the models on the NJR-1, XCorpus, YCorpus, and NYXCorpus datasets}
\label{tab:general-perf}
\begin{tabular}{@{}lc@{\quad}c@{\quad}c@{\quad}c@{\quad}c@{\quad}c@{\quad}c@{\quad}c@{\quad}c@{\quad}c@{\quad}c@{\quad}c@{\quad}c@{\quad}c@{\quad}c@{\quad}c@{}}
\toprule
& \multicolumn{4}{c}{\textbf{NJR-1}} & \multicolumn{4}{c}{\textbf{XCorpus}} & \multicolumn{4}{c}{\textbf{YCorpus}} & \multicolumn{4}{c}{\textbf{NYXCorpus}} \\
\cmidrule(lr){2-5} \cmidrule(lr){6-9} \cmidrule(lr){10-13} \cmidrule(lr){14-17}
{\bf Models} & P & R & F1 & F2 & P & R & F1 & F2 & P & R & F1 & F2 & P & R & F1 & F2 \\
\midrule
\textbf{Random Classifier}  & 0.23 & 0.47 & 0.31 & 0.39 & 0.39 & \textbf{0.48} & 0.43 & 0.46 & 0.22 & 0.45 & 0.29 & 0.37 & 0.25 & 0.47 & 0.33 & 0.40 \\       
\textbf{CGPruner}  & \textbf{0.66} & 0.48 & 0.56 & 0.51 & 0.49 & 0.28 & 0.36 & 0.30 & \textbf{0.71} & 0.24 & 0.36 & 0.28 & 0.61 & 0.43 & 0.50 & 0.46 \\
\textbf{AutoPruner}  & 0.62 & 0.66 & 0.64 & 0.65 & 0.53 & 0.41 & 0.46 & 0.43 & 0.50 & \textbf{0.51} & \textbf{0.50} & \textbf{0.51} & 0.60 & \textbf{0.61} & 0.60 & \textbf{0.60} \\
\textbf{CodeBERT}  & 0.62 & 0.68 & 0.65 & 0.67 & 0.52 & 0.47 & \textbf{0.50} & \textbf{0.48} & 0.50 & 0.48 & 0.49 & 0.49 & 0.59 & 0.60 & 0.60 & \textbf{0.60} \\
\textbf{CodeT5}  & 0.65 & 0.69 & \textbf{0.67} & 0.68 & 0.54 & 0.31 & 0.39 & 0.34 & 0.50 & 0.48 & 0.49 & 0.48 & 0.63 & 0.58 & \textbf{0.61} & 0.59 \\
\textbf{CodeT5+}  & 0.63 & \textbf{0.73} & \textbf{0.67} & \textbf{0.70} & \textbf{0.61} & 0.23 & 0.34 & 0.27 & 0.54 & 0.46 & \textbf{0.50} & 0.48 & \textbf{0.65} & 0.57 & \textbf{0.61} & 0.58 \\
\midrule
\textbf{Average} & 0.57 & 0.62 & 0.58 & 0.60 & 0.51 & 0.36 & 0.41 & 0.38 & 0.49 & 0.44 & 0.44 & 0.43 & 0.55 & 0.54 & 0.54 & 0.54 \\
\midrule
\textbf{Wala} & 0.24 & 0.95 & 0.38 & 0.59 & 0.39 & 0.95 & 0.55 & 0.73 & 0.22 & 0.90 & 0.35 & 0.55 & 0.25 & 0.95 & 0.39 & 0.60 \\
\bottomrule
\end{tabular}
\end{table*}

\paragraph{Datasets characteristics} Table~\ref{tab:data-stats} shows the characteristics of the NJR-1, XCorpus, and YCorpus datasets. NJR-1 has 1.2M samples/edges, whereas XCorpus and YCorpus have 326K and 314K edges, respectively. We also have a "meta" dataset, namely, \textit{NYXCorpus}, by combining the training and test sets of the three datasets. This allows us to compare the performance of the models across all datasets and their programs.

In Table~\ref{tab:data-stats}, we also reported the P/R ratio for each dataset, representing the number of to-be-pruned edges divided by the number of true edges. It can be seen that NJR-1 has a P/R ratio of 18.7, which is much higher than that of XCorpus and YCorpus. This means that the NJR-1 dataset is massively imbalanced. For XCorpus and YCorpus, a lower P/R ratio means that more static edges are observed during test execution at run-time. Also, as expected, NYXCorpus is placed between NJR-1 and Y/XCorpus given its P/R ratio.

\section{Evaluation}\label{sec:eval}
This section presents the motivation, methodology, and empirical results for all research questions that were defined before. 

\subsection{\RQ{1}: \rqOne}
As the first step of our evaluation, we want to explore the overall performance of the different ML models and their capacity to prune static CGs.
This first assessment provides insights into their abilities and allows us to reduce the list to the most promising candidates for the rest of the paper.
 
\paragraph{Methodology} We used our three datasets (NJR-1, XCorpus, and YCorpus) for this experiment.
Specifically, for the NJR-1 dataset, we trained and evaluated the models in the same way prior work did~\cite{utture2022striking}, using 100 programs for training and 41 for evaluation.
For XCorpus/YCorpus, we trained the models on 12/15 programs and evaluated them on 4/3 programs, respectively.
The results also list NYXCorpus, which is a combination of the three datasets that helps us with choosing the overall best-performing models.
For all datasets, Wala's static CGs were constructed using 0-CFA and there is no overlap of programs in the training and test sets. We should also point out that, for the CLMs, we use semantic features if source code is available for an edge/sample. Otherwise, we use the signature-based feature as a fallback.

To find the optimal hyper-parameters for CGPruner (i.e., RandomForest), similar to the work of Utture et al.~\cite{utture2022striking}, we performed 4-fold cross-validation with grid search. In addition to the all models described in subsection~\ref{sec:models}, we also considered a random binary classifier, which prunes/retains edges with an equal probability, i.e., $0.5$. To assess the quality of pruned static call graphs by the ML models, we used the evaluation metrics described in section~\ref{sec:eval-setup}.

\paragraph{Results} Table~\ref{tab:general-perf} shows the general performance of all the models on four datasets, namely, NJR-1, XCorpus, YCorpus, and NYXCorpus.
In addition to the traditional F1 score, we have also included the F2 score in our evaluation, which puts a higher importance on the recall of an approach.
The results show that all the ML models substantially outperform the random classifier across all datasets and all metrics, with the exception of the \emph{recall} in the XCorpus, which seems to be an outlier.
The code language models (i.e., AutoPruner, CodeBERT, and CodeT5(+)) generally perform better than CGPruner at the CG pruning task.
This is expected as the CLMs leverage code semantics whereas CGPruner only relies on structural features.

From Table~\ref{tab:general-perf}, we also observe that all the models perform better on the NJR-1 dataset compared to XCorpus and YCorpus. This is because both XCorpus and YCorpus contain popular real-world Java projects in contrast to NJR-1, which focused on automation over popularity when selecting Java projects~\cite{palsberg2018njr} and popular projects seem to be more difficult for the models. In addition, the ML models perform best on NYXCorpus after NJR-1 with an F2 score of 0.54.
This score shows that the gained precision comes at the price of a reduced recall when compared to Wala.
While the average F2 score is only 0.54 on NYXCorpus, the best models can match the quality of Wala's 0-CFA analysis.
Based on these results, we decided to use CodeBERT and CodeT5 for the subsequent RQs.
These two models perform better than others concerning the F2 score and they do not require structural features, unlike AutoPruner. Also, they are faster compared to CodeT5+ considering both training and inference.

However, these results also bring two follow-up questions.
First, the comparatively low recall and many missing edges can prove impractical for client analyses.
We will explore the effect of more conservative training and pruning strategies in \RQ{2}.
Second, a context-sensitive CG generator might achieve the same performance gain with better soundness.
As such, we will compare the quality and scalability of the results when using a 1-CFA analysis in \RQ{3}.

\NewDocumentCommand{\imgSize}{}{.9\columnwidth}
\NewDocumentCommand{\imgToCap}{}{-14pt}
\NewDocumentCommand{\capToCap}{}{-10pt}
\begin{figure*}
    \hfill
    \begin{subfigure}[b]{\imgSize}
        \centering
        \includegraphics[width=\textwidth]{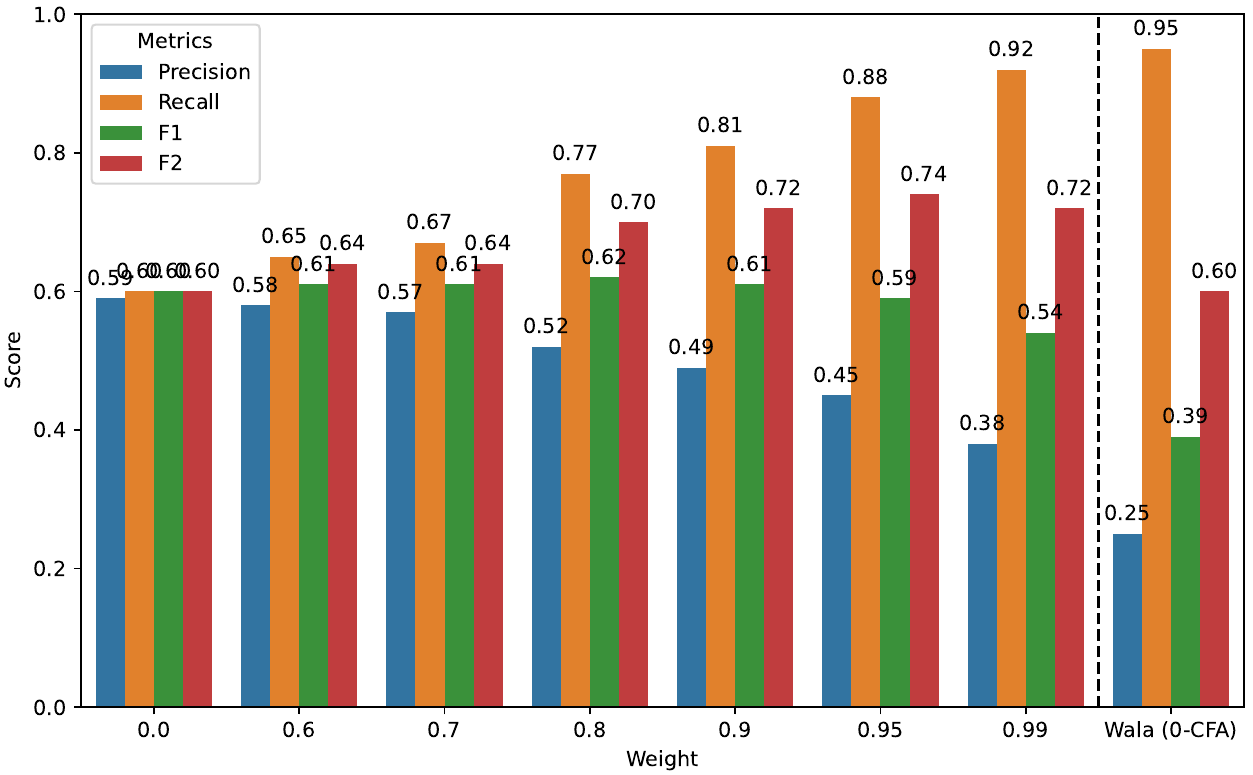}
        \caption{CodeBERT}
        \label{fig:sub1}
    \end{subfigure}
    \hfill
    \begin{subfigure}[b]{\imgSize}
        \centering
        \includegraphics[width=\textwidth]{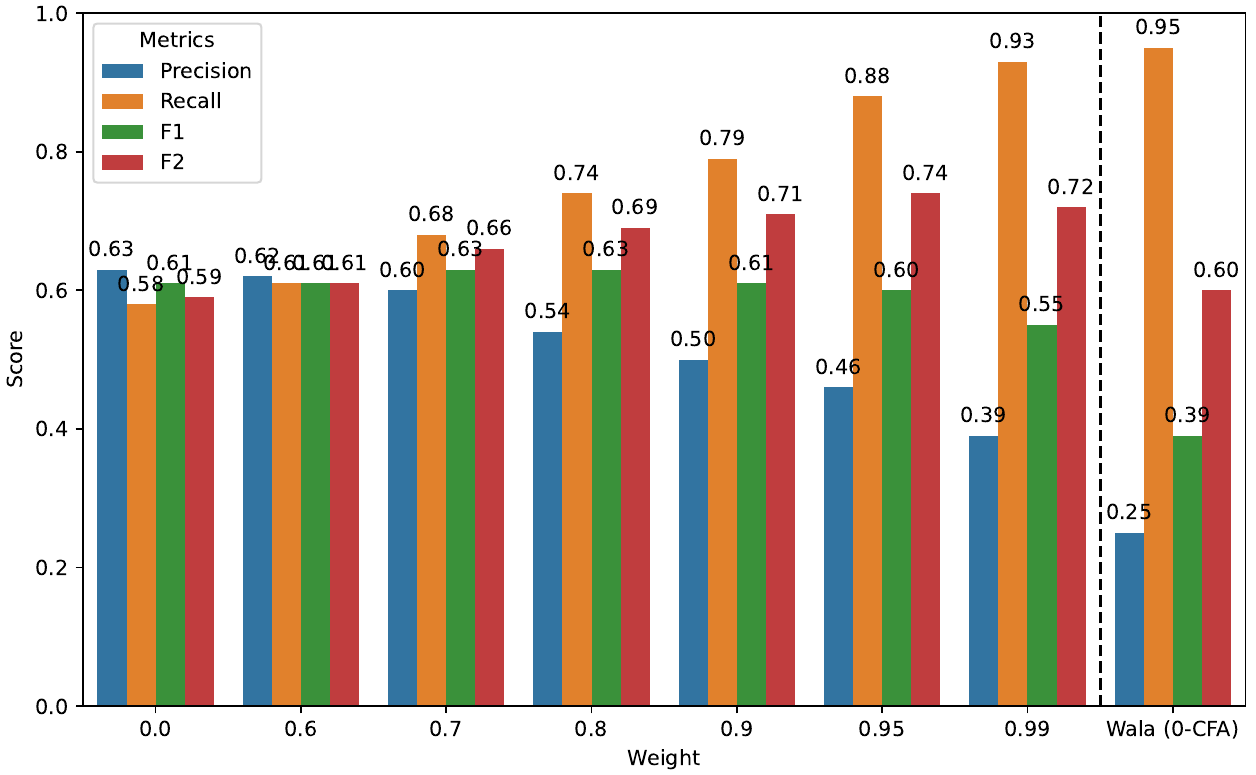}
        \caption{CodeT5}
        \label{fig:sub2}
    \end{subfigure}
    \hfill
    \caption{Performance of the models with different weights to the positive class (retaining edges)}
    \label{fig:weighting-models}
	\hfill
    \begin{subfigure}[b]{\imgSize}
        \centering
        \includegraphics[width=\textwidth]{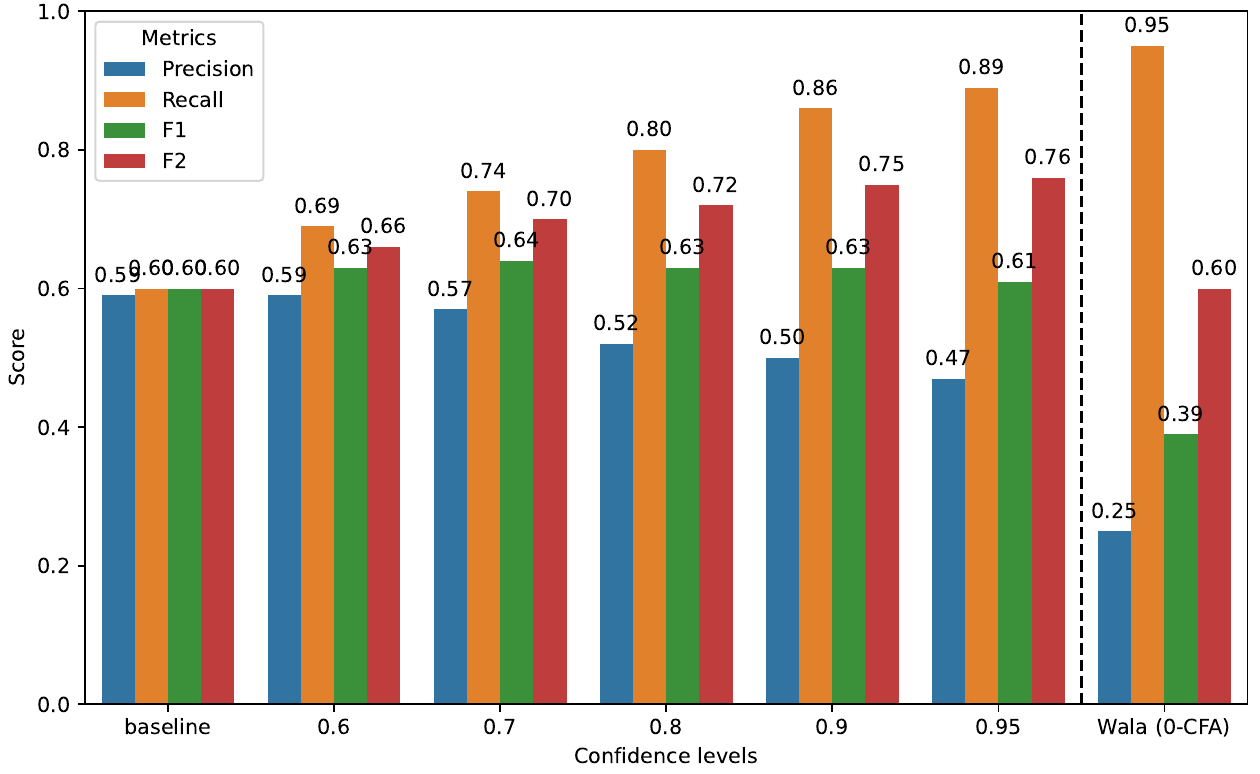}
        \caption{CodeBERT}
        \label{fig:cb-conf-lvl}
    \end{subfigure}
	\hfill
    \begin{subfigure}[b]{\imgSize}
        \centering
        \includegraphics[width=\textwidth]{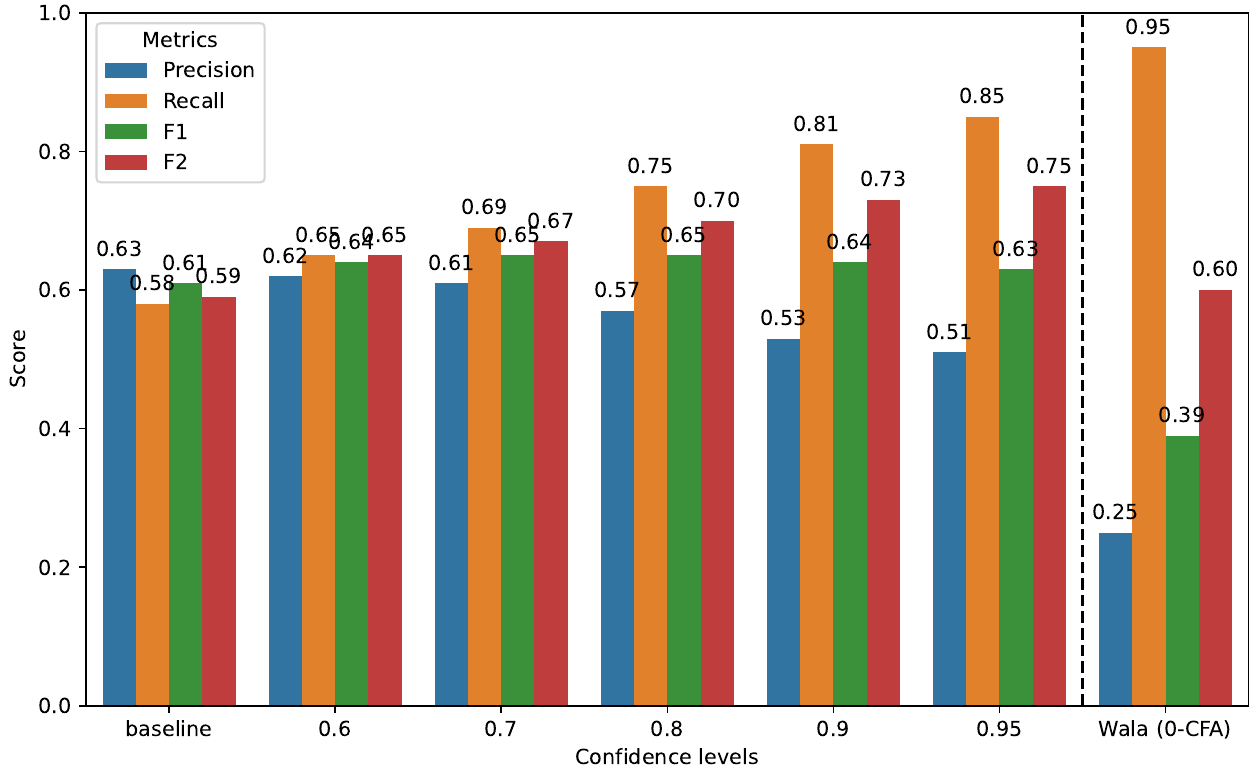}
        \caption{CodeT5}
        \label{fig:models-threshold-t5}
    \end{subfigure}
	\hfill
    \caption{Performance of the models by considering different confidence thresholds}
    \label{fig:models-threshold}
\end{figure*}

\subsection{\RQ{2}: \rqTwo}

The three datasets of our evaluation are imbalanced, especially NJR-1, as can be seen from the P/R ratio in Table~\ref{tab:data-stats}. There are substantially more edges that need to be pruned than edges to be kept.
If both true and false edges are treated equally in the training phase, the ML models will get biased towards pruning.
Indeed, the results of RQ1 have shown that the recall drops significantly, so we will experiment with more conservative strategies during training and pruning to limit the effects of the imbalance.
The goal is to keep as many edges as possible to minimize the false pruning decisions.

\paragraph{Methodology} We use the two most suitable ML-based CGs pruners from \RQ{1}, i.e., CodeBERT and CodeT5, and experiment with two conservative enhancements.
First, we assign a weight $w$ during the learning process to the positive class (i.e., retaining edges) in the cross-entropy loss function (see Equation ~\ref{eq:cross-entropy}) to fine-tune two models separately.
Second, we consider the confidence of the pruning decision and require reaching a configurable threshold $\tau$, before an edge gets pruned.
We investigate the effects in two separate experiments.
In the first experiment, we perform a grid search over the training weights $\{0,6, 0.7, 0.8, 0.9, 0.95, 0.99\}$ to investigate the effect of the weighted loss function.
In the second experiment, we used the unweighted version of the two models that have been used in \RQ{1} already, and performed another grid search over the confidence threshold values $\{0,6, 0.7, 0.8, 0.9, 0.95\}$ in the decision function defined in Equation~\ref{eq:decision_func} to find the best-performing confidence level.
The higher the value of $\tau$, the more conservative we are when pruning CG edges.
Both experiments use 0-CFA-based static CGs.

\paragraph{Results} 
Figure~\ref{fig:weighting-models} shows the performance of the models while fine-tuning them with different weights to the positive and negative classes. For instance, a weight of $0.70$ to the positive class means that the negative class is given a weight of $0.30$. Overall, for both CodeBERT and CodeT5, we observe that the F2 score increases and precision decreases by giving higher weight to the positive class.

The results suggest that the weights of $0.95$ and $0.99$ are the most interesting configurations.
While the F2 score drops slightly when moving from $0.95$ to $0.99$, it is to be expected that the resulting CG is also larger.
It seems that $0.95$ can be considered as the "sweet spot" to achieve a recall close to Wala's while gaining higher precision. The use case for $0.99$ is the application where soundness is essential, like vulnerability analysis.
In short, with weights given to the classes, it is possible to maintain a relatively high recall while having better precision compared to Wala's static call graphs. 

Figure~\ref{fig:models-threshold} indicates the performance of the models considering different confidence levels for pruning call graph edges.
It can be seen that the higher the confidence level, the higher the F2 score is, which is expected as the model is more confident when pruning edges.
Overall, it becomes obvious that the differences to the unweighted results in Figure~\ref{fig:models-threshold} are minimal, and also here, the $0.95$ and $0.99$ levels are the best-performing configurations.
The confidence-based filtering seems to be as effective as using weights in the loss function, while not requiring the additional overhead of fine-tuning.

The similar F2 scores of the original and the pruned CG beg the question of how large the effect of the pruning is in practice.
We will investigate the impact on a client analysis and the performance implication of a substantially reduced CG on runtimes in \RQ{4}.

\subsection{\RQ{3}: \rqThree}

The results of \RQ{1} have shown that ML-based CG pruning can improve a 0-CFA-based CG with a small computational overhead.
The interesting question is how this overhead compares to running more advanced, context-sensitive CG algorithms like k-CFA (i.e., 1-CFA), which has higher precision but is also computationally more expensive.
In this section, we will investigate how using a 1-CFA analysis in the CG generation compares to 0-CFA (with and without pruning) in terms of performance and scalability.

\paragraph{Methodology}
First, we generate static CGs using both 0-CFA and 1-CFA algorithms for the training and test programs in the NYXCorpus dataset.
We reused the CodeBERT and CodeT5 models that have been fine-tuned on the 0-CFA CGs in \RQ{1}.
We also fine-tuned both CLM models on 1-CFA CGs of the NYXCorpus training set with no weight given to the loss function.
When pruning static CGs, we use a confidence threshold of $0.95$, which we found to be an effective configuration in \RQ{2}.

In our second experiment, we measure the CPU time for generating a static call graph using 0-CFA and 1-CFA. In addition, we show how long it takes to prune call graphs by measuring feature extraction and model inference time separately. Lastly, we sum up the CPU time for both static CG generation and CG pruning to show the total computational time of the whole process.
For each measurement, we report the average and the standard deviation across the NYXCorpus programs. 

\paragraph{Results}
Table~\ref{tab:0-1cfa-perf} shows the performance of the CodeBERT and CodeT5 models when fine-tuned on 0-CFA and 1-CFA call graphs for the call graph pruning task. The first observation is that, unsurprisingly, Wala's context-sensitive 1-CFA CGs have a 9\% higher precision than Wala's 0-CFA CGs.
Also, the achieved recall is higher, which in combination results in a substantial increase of both F1 and F2 scores by 11\%.
It is interesting to see that this CG improvement is barely visible in the pruned CGs, which only see a 1-2\% improvement in their F1 and F2 scores.

Table~\ref{tab:run-time-0-1-cfa} provides an overview of the runtimes of the different configurations of 0/1-CFA with and without pruning to allow an assessment of the results in terms of scalability.
CG pruning consists of feature extraction, i.e., tokenizing code sequences and creating semantic features, and model inference, i.e., querying the CLM model to prune call graph edges.
Clearly, the CG pruning task adds additional computational overhead on top of the static CG generation.
The table therefore splits the different stages and shows the averages and standard deviations for static CG generation, feature extraction and inference, and the total runtime in seconds.

The results show that a 1-CFA-based static CG generation takes 42.3s, which is almost twice as long as the 0-CFA algorithm without pruning (21.4s).
The 1-CFA alternative is therefore as expensive as 0-CFA with pruning ($\approx 42$s), however, its standard deviation is much higher (120s vs. 65s).
This is likely caused by the computational complexity of static analysis, which, unlike ML models that have a constant runtime per query, does not scale linearly with the program size.
The results suggest that context-sensitive analysis can be beneficial for small programs, while ML-based approaches scale better.
It is worth noting that the improved CGs of the 1-CFA analysis also have a positive impact on the runtime of the ML approaches.
As the CGs are smaller and contain fewer edges, the runtime of the pruner goes down and deviates less.

\begin{table}[t]
\centering
\caption{Model Performance with 0/1-CFA algorithm}
\label{tab:0-1cfa-perf}
\begin{tabular}{@{}l@{\quad}c@{\quad}c@{\quad}c@{\quad}c@{}}
\toprule
\multirow{2}{*}{\textbf{Models}} & \multicolumn{4}{c}{\textbf{NYXCorpus}} \\
\cmidrule(lr){2-5}
 & P & R & F1 & F2 \\
\midrule
\multicolumn{5}{@{}l}{\it 0-CFA} \\ 
\midrule
\textbf{CodeBERT}  & 0.47 & 0.89 & 0.61 & \textbf{0.76} \\                                                                
\textbf{CodeT5}  & \textbf{0.51} & 0.85 & \textbf{0.63} & 0.75 \\
\textbf{Wala} & 0.25 & {\bf 0.95} & 0.39 & 0.60 \\
\midrule
\multicolumn{5}{@{}l}{\it 1-CFA} \\ 
\midrule
\textbf{CodeBERT}  & 0.49 & 0.91 & 0.64 & \textbf{0.78} \\                                                                
\textbf{CodeT5}  & \textbf{0.53} & 0.86 & \textbf{0.65} & 0.76 \\
\textbf{Wala} & 0.34 & {\bf 0.97} & 0.50 & 0.71 \\
\bottomrule
\end{tabular}
\end{table}

\begin{table}[t]
\caption{Runtime of 0/1-CFA algorithms with CG pruning}
\label{tab:run-time-0-1-cfa}
\resizebox{\columnwidth}{!}{%
\begin{tabular}{@{}lccccc@{}}
\toprule
\multirow{2}{*}{\textbf{Models}} & \multirow{2}{*}{\textbf{CG Gen. [s]}} & \multicolumn{2}{c}{\textbf{Pruning [s]}} & \multirow{2}{*}{\textbf{Total Time [s]}} \\
\cmidrule(lr){3-4}
&    & Feature & Infer. &    \\ 
\midrule
\multicolumn{5}{@{}l}{\it 0-CFA} \\ 
\midrule
\textbf{CodeBERT} & \multirow{2}{*}{\mypm{21.4}{57.0}} & \multirow{2}{*}{\mypm{2.8}{3.5}} & \mypm{18.6}{30.9} & \mypm{42.7}{65.2}\\
\textbf{CodeT5} & & & \mypm{18.9}{31.4} & \mypm{43.0}{65.4} \\
\midrule
\multicolumn{5}{@{}l}{\it 1-CFA} \\ 
\midrule
\textbf{CodeBERT} & \multirow{2}{*}{\mypm{42.3}{120.5}} & \multirow{2}{*}{\mypm{1.4}{0.8}} & \mypm{11.7}{11.2} & \mypm{55.1}{122.6} \\
\textbf{CodeT5} & & & \mypm{14.5}{20.0} & \mypm{57.9}{123.9} \\
\bottomrule
\end{tabular}
}
\end{table}

\subsection{\RQ{4}: \rqFour}

All previous experiments have used statistical means to assess the pruning performance of the ML models by comparing ground truth and pruned CG through metrics such as F2-score.
Previous works have employed client analyses on the pruned CGs, like null pointer exceptions (NPE), to show the effects of the pruning on static analyses in practice~\cite{utture2022striking, le2022autopruner}.
We follow this example and study the effects of CG pruning on vulnerability propagation, a security-sensitive analysis that requires the traversal of call graphs~\cite{mir2023effect}.
We will report on the resulting CG sizes and the runtime of the client analysis.
We expect a significant speed-up in finding vulnerable call paths using pruned static call graphs, though pruned CGs may be susceptible to false negatives, which needs to be investigated.

\paragraph{Methodology} 
We used WALAs 0-CFA static CGs as the baseline and we employ the CodeBERT and CodeT5 models that have been fine-tuned on the training set of NYXCorpus.
We use two configurations for the comparison.
The \emph{conservative} configuration does not add weights to the loss function but uses a $0.95$ confidence threshold.
The \emph{paranoid} configuration reuses the CodeBERT and CodeT5 models that have been fine-tuned with a weight of $0.99$ to the positive class and applies a confidence level of $>0.95$ for pruning.

Our experimental design builds upon the availability of method-level vulnerability information, which is provided by tools like Prospector~\cite{ponta2019msr}.
We did not use real-world vulnerabilities for programs in NYXCorpus though, as the NJR-1 dataset does not include Maven coordinates for its dependencies, plus many projects without vulnerabilities would have to be filtered.
As such, we decided to randomly mark 100 methods as vulnerable in each program of the NYXCorpus test set.
All marked CG nodes represent non-application nodes that are defined in the dependencies of a program.
Our experimental goal is then to measure how long it takes to compute all paths that start in the application and reach a vulnerability with a simple reachability analysis through a Breadth-first-search (BFS).
We calculate the fraction of vulnerable methods that are reachable in a given CG, before and after pruning.
Also, there is no reason to believe that our artificial vulnerabilities are easier to reach than actual vulnerabilities.
To accurately measure the analysis time, we first run the code three times to warm up the JVM and let the JIT compilation do its optimization.
We then compute the reachability another three times and average the required execution time.

It is worth pointing out that while the absolute number of identified paths is lower in a pruned CG, we believe that the crucial information is whether a vulnerable method is reached at all.
Moreover, it is irrelevant for the actionability of the results, whether 1 or 10 affected paths can be found for a given vulnerability.

\begin{table}
\caption{Vulnerability Propagation Analysis on (pruned) CGs}
\label{tab:vuln-analysis-results}
\begin{threeparttable}
\resizebox{\columnwidth}{!}{%
\begin{tabular}{@{}lrrrrr@{}} 
\toprule
\multirow{2}{*}{\bf Models} & \multicolumn{2}{c}{\bf CG Size} & \multicolumn{2}{c}{\bf Reachable ...} & \multirow{2}{*}{\bf Time (ms)} \\
\cmidrule(lr){2-3} \cmidrule(lr){4-5}
 & Edges& Nodes\tnote{1} & .. Paths & .. Nodes & \\
\midrule
\textbf{Wala} & 5227.1 & 1420.2 & 853.9 & 99.8\% & \mypm{8.1}{26.6} \\
\midrule
\multicolumn{6}{@{}c@{}}{{\it Conservative Pruning} \hfill ($>0.95$ confidence)} \\ 
\midrule
\textbf{CodeBERT} & 1736.4 & 1048.2 & 515.3 & 86.0\% & \mypm{2.3}{8.1} \\
\textbf{CodeT5} & 1498.2 & 950.5 & 388.6 & 82.0\% & \mypm{1.5}{4.8} \\
\midrule
\multicolumn{6}{@{}c@{}}{{\it Paranoid Pruning} \hfill ($0.99$ weight, $>0.95$ confidence)} \\ 
\midrule
\textbf{CodeBERT} & 3728.4 & 1392.2 & 778.9 & 98.4\% & \mypm{6.6}{23.0} \\
\textbf{CodeT5} & 3503.5 & 1337.3 & 832.5 & 96.9\% & \mypm{6.7}{23.0} \\
\bottomrule
\end{tabular}
}
\begin{tablenotes}
    \item[1] \#Nodes with at least an incoming and/or out-going edge
\end{tablenotes}
\end{threeparttable}

\end{table}

\paragraph{Results}  
Table~\ref{tab:vuln-analysis-results} shows the results of vulnerability propagation for both 0-CFA-based static CGs and their pruned version. Note that the reported numbers for CG size, the number of reachable vulnerable paths and nodes are average per test program in NYXCorpus.

WALA is the baseline for the comparison.
It is obvious that both pruning strategies are able to substantially reduce the original CG size from 5.2K edges to 1.5-1.7K ($\approx 33$\%) with the \emph{conservative} setting and 3.5-3.7K ($\approx 69$\%) in the \emph{paranoid} setting.
This results in substantial reductions in the runtime of the client analysis to only 1.5ms (5.4x speedup) in CodeT5 and 2.3ms (3.5x speedup) in CodeBERT.
While the concrete reachability analysis is very fast even on the original CG, we have already seen earlier in the paper that static analyses scale non-linearly, so every reduction in the size of a CG will have a substantial impact in more advanced analyses. 

The substantial reduction of the \emph{conservative} setup comes at the price of only reaching 86\% of the vulnerable nodes.
However, the \emph{paranoid} setup is able to retain the reachability of 96-98\% of the vulnerable nodes, which comes very close to the WALA baseline.
We find that the reduced size and substantial speedup make this result very attractive for large-scale analyses, but the best CG choice always depends on the task and the context.
A security-focused application might accept the slower execution time of an unpruned 1-CFA-based CG analysis to gain a sound result.
For other use cases, a \emph{paranoid} setup might be all that is required, or even a \emph{conservative} analysis could work, when performance is the main issue.
It is noteworthy that, at least in the presented analysis, the pruning does not introduce any false positives and only introduces a small fraction of false negatives. 

\section{Discussion}\label{sec:discuss}
When reflecting on the obtained evaluation results, we believe that several points are noteworthy and should be considered by researchers and practitioners.

\paragraph{Call graph pruning is an open problem}
As shown throughout the paper, code language models like CodeBERT and CodeT5 have the potential to substantially improve the precision of static CGs.
However, we have also seen that CG pruning is challenging, especially for the real-world programs in the XCorpus and YCorpus.
While the precision is good, the main challenge is achieving a reasonably good recall as well.
The probabilistic nature of CLM models makes it easy to introduce pruning thresholds, however, the parameters of our current approaches must be fine-tuned in a small range at the extremes.
Our models can present a promising step in the right direction, but they do not give an exhaustive answer to the larger problem.
More work is required to find a more robust approach with a more differentiated confidence measure and better results overall.
Future work could explore hybrid approaches that combine heuristic (non-) pruning rules with a CLM model.

\paragraph{Data Imbalance}
We believe that the main limitation that we have faced in our experiments is the massive imbalance of the dataset, as seen in the P/R ratio in Table~\ref{tab:data-stats}).
Naturally, trained ML models will be biased towards pruning edges rather than retaining them.
We believe that future work should continue to emphasize \emph{recall} over \emph{precision}, as we have done by using the \emph{F2} measure when optimizing their models.
However, this is only the first step and further approaches need to be taken to counter the imbalance.
Our technique for building the ground truth was executing test suites for collecting relevant edges.
Future work could extend this endeavor and trace more extensive program executions and build more complete dynamic CGs.

\paragraph{Hybrid Static Analysis}
Recent works have introduced advancements in ML-based CG pruning, but also advanced program analysis approaches that consider call site sensitivity and more context to improve the precision of static CGs~\cite{ma2023context, jeon2022return}.
Unfortunately, the ML-based approaches and advanced static analyses are still often seen as related, but separate solutions to the CG generation problem.
Likely, because both are very advanced topics in their respective fields and because it is hard to find researchers who are experts in both areas.
We strongly believe though that a hybrid static analysis that integrates ML-based approaches into the static analysis instead of running it as a separate step would be very promising.
Our experiments have shown that even the best static CG generator that we included could not reach 100\% of the vulnerable nodes.
This is, for example, caused by calls through the Java reflection API, which could be suggested through complementing probabilistic approaches.
Another potential combination would be the use of ML to improve the performance of advanced static analyses, which would otherwise not scale to large programs, for example, by accepting unsound results for less important parts of the program.

\paragraph{Feature Engineering}
In contrast to mostly structural features and graph metrics that have been used in previous work, we used semantic features that are based on the source code that surrounds the potential call site.
Overall, we believe that the obtained results are positive, but the feature engineering idea should be investigated more closely.
It is likely that considering full methods for sources and targets exceeds the attention of ML models, therefore important information is not taken into consideration by the model.
Future work could investigate new ways to encode the surrounding context of a method call to find better, semantic features, which might carry more relevant information about the likelihood of a call relation between two methods.

\section{Threats to Validity and Limitations}\label{sec:ttv}
In this section, we describe the limitations of our work, possible threats to the validity of our empirical findings, and how we address them. We have picked F2 as the main metric to judge the CG quality of our pruned CGs. While it could be possible that the metric still does not emphasize the importance of recall enough, we believe that our results in the vulnerability analysis confirm the suitability of the metric in our experiments.

Our experimental result rests on the ground truth that we have generated through the instrumentation and execution of test suites, which might not be complete or representative of other programs.
We selected programs with a high test coverage, which makes us confident that the results are reliable. Larger benchmark datasets will certainly contain more cases that might be missed in this paper, but they would also provide more data to train the ML models.
Overall, we are confident about the representativeness of our data, confirming the data with larger datasets will be left for future work.

We have chosen a vulnerability analysis as a client analysis that is built upon a CG. We do not even start to object that this choice might not be representative of other analysis tasks. However, we think that the generated results and the insights still hold, as the described downsides of static analyses only get worse with larger programs or more advanced static analysis algorithms.

Lastly, we filter call graph edges based on package names as described in subsection~\ref{subsec:CGG}. This may cause the exclusion of call graph edges related to Graphical User Interface (GUI) components or event-driven programming aspects from the evaluation. This is not a threat but a limitation of the filtering strategy we used, following the previous work~\cite{le2022autopruner, utture2022striking}, which filters such edges if their CG node's URI starts with any of those filtered packages like \texttt{java/}. Future work should propose a more robust approach to filtering call graph edges before the training phase.

\section{Summary}\label{sec:summary}
This paper presents an empirical study on the effectiveness of machine learning-based call graph pruning. We identified several key issues in the current state of research on ML-based call graph pruning such as a lack of a suitable benchmark dataset, data imbalance due to static analysis over-approximation, significant recall drop in CG pruning, and no comparison between pruned 0-CFA-based call graphs with context-sensitive algorithms like $k$-CFA.
To address these challenges, we have introduced
(1) the NYXCorpus dataset, combining NJR-1, XCorpus, and YCorpus.
(2) and a conservative strategy to prune CG edges more confidently, which can be tuned by giving weights to classes in the learning process or considering different confidence levels when pruning.
Our empirical findings show substantial improvement in CG precision.
Specifically, ML-based CG pruning can boost precision by 24-34\% while reducing the recall by 2-10\%.
Even though our experiments favor recall over precision, we can show through a comparison with a more advanced 1-CFA-based CG generation that the overall tradeoff is in favor of the ML-based approaches.
We show in a client analysis that by tweaking our model parameters to a \emph{paranoid} setup, it is possible to achieve virtually identical results to a static analysis while being 3.5x faster and operating on a reduced CG with 69\% of its original size.


\section*{Data Availability}
The datasets including NYXCorpus, all the fine-tuned CLMs, and the source code to replicate the results of the experiments in this paper are publicly available. \\
Data \& Models:  \href{https://zenodo.org/doi/10.5281/zenodo.10638852}{https://zenodo.org/doi/10.5281/zenodo.10638852} \\
Source code: \href{https://github.com/mir-am/ml4cgp\_study}{https://github.com/mir-am/ml4cgp\_study}

\begin{acks}
The FASTEN project has received funding from the European
Union’s Horizon 2020 research and innovation program
under grant agreement number 825328. Also, we would like to thank the anonymous reviewers and Georgios Gousios for providing valuable feedback to improve this paper.
\end{acks}

\bibliographystyle{ACM-Reference-Format}
\balance
\bibliography{main}


\end{document}